\documentclass[conference]{IEEEtran}
\IEEEoverridecommandlockouts

\usepackage{cite}
\usepackage{amsmath,amssymb,amsfonts}
\usepackage{algorithmic}
\usepackage{graphicx}
\usepackage{textcomp}
\usepackage{xcolor}
\usepackage{hyperref}
\usepackage{color}
\usepackage{tabularx}
\usepackage{booktabs}
\usepackage{multirow}
\usepackage{url}
\usepackage{float}
\usepackage{makecell}
\usepackage[subrefformat=parens,labelformat=parens]{subfig}
\usepackage{calrsfs}
\DeclareMathAlphabet{\pazocal}{OMS}{zplm}{m}{n}

\makeatletter
\renewcommand\IEEEkeywordsname{Keywords}
\makeatother
\usepackage{xpatch}
\xpatchcmd\IEEEkeywords{---}{-}{}{}

\makeatletter
\renewcommand{\fnum@figure}{Figure~\thefigure}
\makeatother

\usepackage{caption}
\def\BibTeX{{\rm B\kern-.05em{\sc i\kern-.025em b}\kern-.08em
    T\kern-.1667em\lower.7ex\hbox{E}\kern-.125emX}}

\begin{document}

\title{{\bfseries\Large InterGridNet: An Electric Network Frequency Approach for Audio Source Location Classification Using Convolutional Neural Networks}\\

\author{Christos Korgialas\IEEEauthorrefmark{1}, \IEEEauthorblockN{Ioannis Tsingalis\IEEEauthorrefmark{1}, Georgios Tzolopoulos\IEEEauthorrefmark{1}, and Constantine Kotropoulos}
\IEEEauthorblockA{\textit{Department of Informatics} \\
\textit{Aristotle University of Thessaloniki}\\
Thessaloniki 54124, Greece\\
email: \{ckorgial, tsingalis, gtzolopo, costas\}@csd.auth.gr}
\IEEEauthorblockA{\IEEEauthorrefmark{1} Equal contribution}}}

\maketitle

\begin{abstract}
A novel framework, called InterGridNet, is introduced, leveraging a shallow RawNet model for geolocation classification of Electric Network Frequency (ENF) signatures in the SP Cup 2016 dataset. During data preparation, recordings are sorted into audio and power groups based on inherent characteristics, further divided into 50 Hz and 60 Hz groups via spectrogram analysis. Residual blocks within the classification model extract frame-level embeddings, aiding decision-making through softmax activation. The topology and the hyperparameters of the shallow RawNet are optimized using a Neural Architecture Search. The overall accuracy of InterGridNet in the test recordings is 92\%, indicating its effectiveness against the state-of-the-art methods tested in the SP Cup 2016. These findings underscore InterGridNet's effectiveness in accurately classifying audio recordings from diverse power grids, advancing state-of-the-art geolocation estimation methods.
\end{abstract}

\begin{IEEEkeywords}
\textit{electric network frequency (ENF); grid location estimation; audio processing; multimedia forensics}
\end{IEEEkeywords}

\section{Introduction}
\label{sec:intro}

Due to power grid disturbances, the Electric Network Frequency (ENF) is a dynamic time series that exhibits fluctuations around its nominal frequency of 50 Hz in Europe and 60 Hz in the United States/Canada. These oscillations result from instantaneous load variations within the power grid, displaying a consistent pattern within interconnected grids. ENF signals become embedded in multimedia recordings captured in proximity to power sources. This distinctive signal can subsequently be extracted \cite{vatansever2019,korgialas2024,korgialas2024robust,moysiadis2023electric} from digital recordings for various applications, such as verification of recording timestamps~\cite{garg2012modeling,hua2018,zheng2017,hua2014}. 

Another application where ENF is also utilized is grid localization. Grid localization can be treated as inter-grid~\cite{hajj2013,hajj2015,tzolopoulos2024} or intra-grid~\cite{hajj2015}\cite{yao2017source}\cite{garg2021} localization. Inter-grid localization focuses on identifying the grid in which a media recording was captured, while intra-grid localization aims to determine the recording's location within the specific grid precisely. Inter-grid localization is briefly surveyed in Section~\ref{sec:relatedWork}. 

The intra-grid localization is considered more challenging due to the highly subtle distinctions in the ENF signatures encoded within the recorded signals. However, this assumption is challenged by \cite{garg2013geo}, who detail noticeable differences due to varying city power consumptions and the time for load changes to impact different grid segments, a concept further explored in \cite{elmesalawy2014}. Additionally, ENF fluctuations can stem from system disruptions like short circuits, power line switching, and generator failures, as noted in \cite{bollen2006}. Minor local load changes affect ENF differently than major events like generator disconnections, which impact the entire grid at about 500 miles per second \cite{tsai2007}. Given the aforementioned intra-grid characteristics, various methods have been proposed to tackle the problem of intra-grid localization~\cite{hajj2012}\cite{jeon2018m}\cite{cui2018}. 

A novel framework, termed InterGridNet, is introduced for geolocation classification exploiting the ENF. The framework offers a comprehensive approach that includes data preparation and preprocessing techniques using a shallow RawNet \cite{jung2019rawnet} for classification. The topology and the hyperparameters of InterGridNet are optimized through Neural Architecture Search (NAS), enhancing its capability to tackle inter-grid localization in audio recordings. It incorporates innovative techniques, including filtering to isolate the relevant ENF signal, using residual layers to extract frame-level embeddings, and employing a softmax activation function in the decision-making process. To our knowledge, this represents a pioneering advancement spanning from preprocessing techniques to the classification stage, establishing a novel framework in geolocation classification using deep learning methodologies. The Signal Processing (SP) Cup 2016 dataset \cite{sp2016}, the only benchmark dataset publicly available, is employed for assessing geolocation classification.

The key contributions of the paper are as follows:
\begin{itemize}
    \item A novel framework, coined InterGridNet, is proposed to treat geolocation estimation as a classification problem among nine power grids, employing a shallow RawNet optimized with NAS and leveraging ENF signatures from the benchmark SP Cup 2016 dataset. It should be noted that a shallow RawNet is utilized to reduce the number of parameters and achieve comparable performance with that using a deeper neural network.
    \item Experimental evaluation, including extensive testing of the SP Cup 2016 dataset, showcases the effectiveness of InterGridNet in geolocation classification across nine distinct power grids, where it is compared with state-of-the-art methods.
\end{itemize}

The remainder of the paper is organized as follows. Section~\ref{sec:relatedWork} provides an overview of related work. The proposed framework is detailed in Section~\ref{sed:framework}. Experimental evaluation is conducted in Section~\ref{sec:experiments}. Section~\ref{sec:conclusions} concludes the paper by providing information for future work.

\section{Related Work} \label{sec:relatedWork}

ENF variations due to load fluctuations and grid frequency control help to localize audio recordings. Grigoras's research demonstrated this by correlating ENF from audio recordings with reference ENF signals from different power grids to estimate the location of the recording \cite{grigoras2005digital}. Extensive research was conducted in grid localization using ENF by employing diverse datasets~\cite{yao2017source}. Additionally, location estimation at various scales was addressed in \cite{sarkar2019application} and \cite{garg2021}. In~\cite{hajj2015}, a machine learning system was developed to ascertain where an ENF-containing media file was recorded, even when no simultaneous ENF reference was available. Five machine learning algorithms were explored to identify the recording location of power and audio recordings obtained from ten distinct power grids in~\cite{vsaric2016improving}. The hypothesis that variations in load conditions could generate unique location-specific patterns within the ENF signal was assessed in~\cite{garg2013geo}. In \cite{li2024advanced}, an ENF region classification model, UniTS-SinSpec, was introduced within the UniTS framework, integrating a sinusoidal activation function and a spectral attention mechanism and trained on a public dataset. Addressing the complexities of inter-grid classification, field specialists have formulated methodologies to distinguish audio recordings across global power grids, exemplified by the 2016 SP Cup. This work substantially improved the forensic analysis based on ENF, fortifying the verification of the authenticity of multimedia recordings. These distinctive patterns could pinpoint the precise location within a grid where the recording was made.

\section{The InterGridNet Framework}\label{sed:framework}

\begin{figure*}[!ht]
    \centering
    \includegraphics[width=0.8\textwidth]{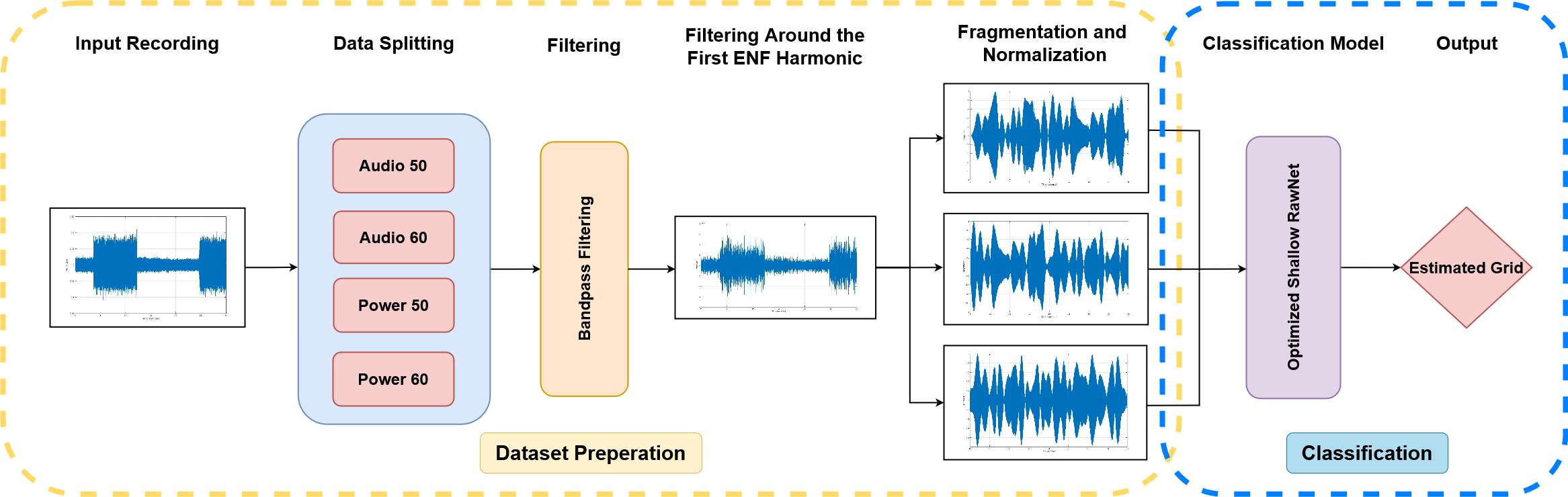} 
    \caption{Flowchart of the proposed InterGridNet framework.}
    \label{fig:flowchart}
\end{figure*}

\subsection{Dataset Preperation}\label{subsec:Dataset}

The SP Cup 2016 competition~\cite{sp2016} benchmark dataset~\cite{SPCup2016} is employed, with data split into three sets: a training set for the model's development and training, a practice set for validation, and a testing set for evaluating performance on unseen data (see Section~\ref{subsec:testing}). The dataset encompasses audio recordings capturing ENF signals from power grids across different countries. Specifically, it consists of recordings from nine distinct power grids, denoted as \textbf{A} through \textbf{I}. Grids \textbf{A}, \textbf{C}, and \textbf{I} are characterized by nominal ENF at 60Hz, while the remaining grids exhibit ENF around 50Hz.

The dataset consists of audio and power recordings for each grid. The power recordings were generated from a specialized circuit designed to capture the ENF time series directly from the power mains and have a temporal span of 30 to 60 minutes. The audio recordings were acquired using a microphone near substantial electrical devices, capturing the ENF hum for 30 minutes. In particular, power recordings are distinguished by stronger ENF traces than audio recordings.

The testing set has been augmented with 100 samples (40 Audio and 60 Power), each spanning 10 minutes. This subset comprises 8-11 samples from each of the nine grids (\textbf{A} - \textbf{I}) and  10 additional samples from networks other than these, categorized as ``None'' (\textbf{N}). This diverse sample set is a benchmark for assessing the proposed InterGridNet's efficiency and generalization. 

Figure~\ref{fig:flowchart} illustrates the InterGridNet framework, highlighting the two critical stages of data preparation and classification. The data preparation process is depicted within the yellow dashed box in Figure~\ref{fig:flowchart}. Initially, the recordings' inherent characteristics, encompassing ENF signals at either 50Hz or 60Hz, are utilized to classify the recordings as audio or power recordings. Four distinct and independent data groups were created: \texttt{audio50}, \texttt{audio60}, \texttt{power50}, and \texttt{power60}. 

During the training phase, this categorization is direct since the differences between audio and power recordings are perceptible, mainly due to the higher Signal-to-Noise Ratio (SNR) in power recordings. In contrast, during the testing phase, an automated grouping method is required to classify recordings based on their spectral characteristics, mainly to distinguish between the ENF frequencies of 50Hz and 60Hz. This method is described as follows:

\begin{enumerate}
    \item For each recording, the average spectrogram magnitude is calculated for the first three harmonics associated with the nominal frequencies of 50 Hz and 60 Hz.
    \item For each nominal ENF, the harmonic with the smallest value from step 1 is ignored. Since the ENF may not be present in every harmonic, the two harmonics with the stronger traces are enough for the categorization.
    \item The average of the magnitude values at the retained frequencies in step 2 is calculated.
    \item The largest value reveals the nominal frequency of the network.
\end{enumerate}

\begin{figure}[t]
    \centering
    \includegraphics[width=0.35\textwidth]{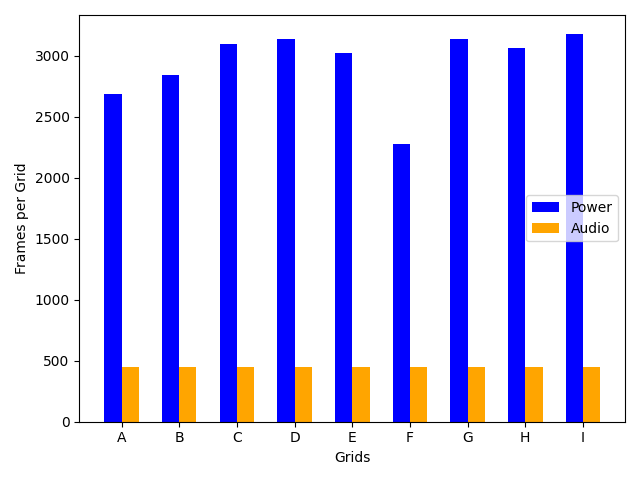}
    \caption{Number of audio and power recording frames in each grid.}
    \label{fig:samples}
\end{figure}

After classifying each recording into its data group, a filtering process is applied to isolate the corresponding ENF within a range of 2 Hz. For instance, samples from the \texttt{audio60} group undergo filtering using a bandpass filter set to frequencies between 59 Hz and 61 Hz. Subsequently, the waveforms are segmented into 16-second frames with a 50\% overlap and normalized to the interval $[-1, 1]$. These processed frames are subsequently fed into the classification model for power grid classification,  shown as the blue dashed box in Figure~\ref{fig:flowchart}. The aggregated count of frames for each grid is depicted in Figure~\ref{fig:samples}, providing an overview of the distribution of frames across the dataset.

\subsection{Classification Architecture}\label{Se:DescAlgo}

\begin{figure*}[!ht]
    \centering
    \subfloat[Grid A (ENF 60Hz).\label{fig:spectroA}]{
        \includegraphics[width=0.23\linewidth]{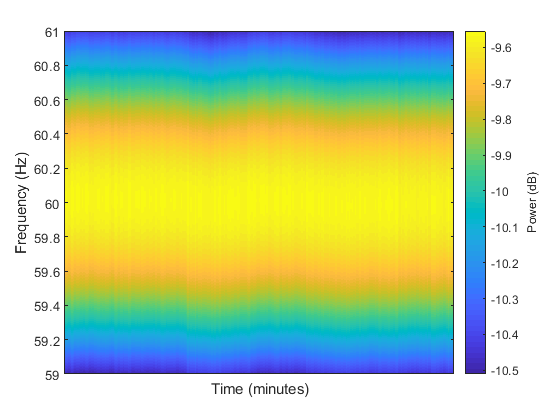}
    }
    \subfloat[Grid B (ENF 50Hz).\label{fig:spectroB}]{
        \includegraphics[width=0.23\linewidth]{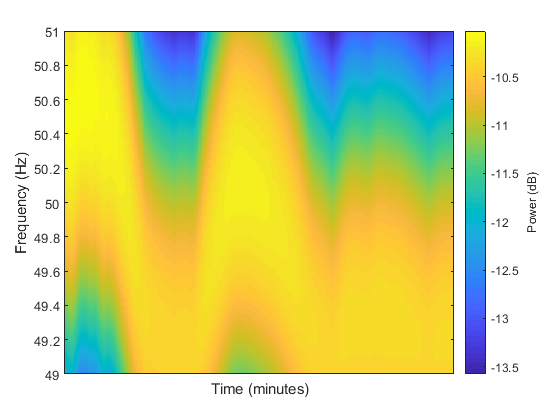}
    }
    \subfloat[Grid D (ENF 50Hz).\label{fig:spectroD}]{
        \includegraphics[width=0.23\linewidth]{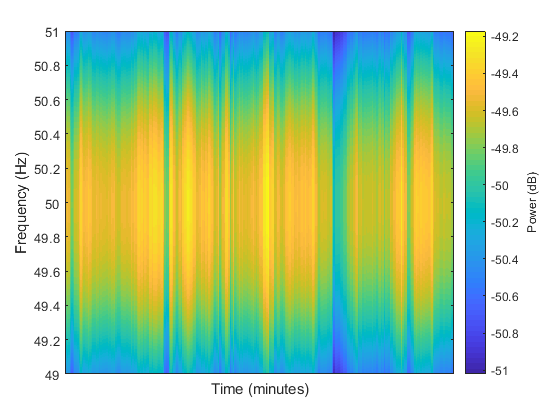}
    }
    \subfloat[Grid I (ENF 60 Hz).\label{fig:spectroI}]{
        \includegraphics[width=0.23\linewidth]{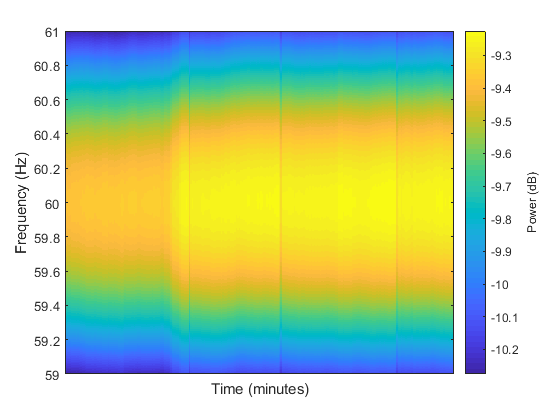}
    }
    \caption{Spectrograms focused on the nominal ENF value for different grids.}
    \label{fig:compositeSpectograms}
\end{figure*}

The spectral content of the frames exhibits variation based on the grid of origin, providing valuable information for the location estimation of the recording. Figure~\ref{fig:compositeSpectograms} displays a spectrogram concentrated around the nominal ENF from four distinct grids. Notably, the ENF behavior differs depending on the grid, wherein Figures~\subref*{fig:spectroA},~\subref*{fig:spectroB},~\subref*{fig:spectroD}, and~\subref*{fig:spectroI} the frequency content is around 60Hz, 50Hz, 50Hz, and 60Hz, respectively. Consequently, the technique elucidated following harnesses these ENF characteristics to classify the samples according to the grid where the recording was made.

To address the classification problem, individual classes are defined for each grid, comprising 16-second frames derived in Section~\ref{subsec:Dataset}. These 16-second frames are called samples hereafter. The classification problem for each data group is denoted as $G^{\mathrm{ENF}}_{\mathrm{REC}}$, where $REC$ represents the recording type (Audio or Power), and $ENF$ signifies the nominal frequency of the grid. Consequently, the classification problems are denoted as $G^{\mathrm{50}}_{\mathrm{Audio}}$, $G^{\mathrm{50}}_{\mathrm{Power}}$, $G^{\mathrm{60}}_{\mathrm{Audio}}$, and $G^{\mathrm{60}}_{\mathrm{Power}}$. 
Each $G^{\mathrm{ENF}}_{\mathrm{REC}}$ is expressed as $G^{\mathrm{ENF}}_{\mathrm{REC}} = \{C_1, C_2, \dots, C_n\}$, where $n=3$ for $G^{\mathrm{60}}_{\mathrm{Audio}}$ and $G^{\mathrm{60}}_{\mathrm{Power}}$, and $n=6$ for the others. Each $C_i$ class in the classification problem contains all samples from the corresponding grid in the respective data group.

\begin{table}[t]
    \caption{OPTIMIZED HYPERPARAMETERS OF THE SHALLOW RAWNET.}
    \label{table:hyper}
    \begin{center}
    \scalebox{0.9}{
    \begin{tabular}{ccccc}
        \toprule
        & $G^{\mathrm{50}}_{\mathrm{Audio}}$ & $G^{\mathrm{60}}_{\mathrm{Audio}}$ & $G^{\mathrm{50}}_{\mathrm{Power}}$ & $G^{\mathrm{60}}_{\mathrm{Power}}$ \\
        \midrule
        Learning Rate & $6.5 \times 10^{-4}$ & $7 \times 10^{-4}$ & $1.1 \times 10^{-3}$ & $9.7 \times 10^{-4}$ \\
        $\beta_1$ & 0.96 & 0.97 & 0.98 & 0.98 \\
        $\beta_2$ & 0.998 & 0.998 & 0.992 & 0.993 \\
        \bottomrule
    \end{tabular}}
    \end{center}
\end{table}

\begin{figure*}[t]
    \centering
    \includegraphics[width=0.85\textwidth]{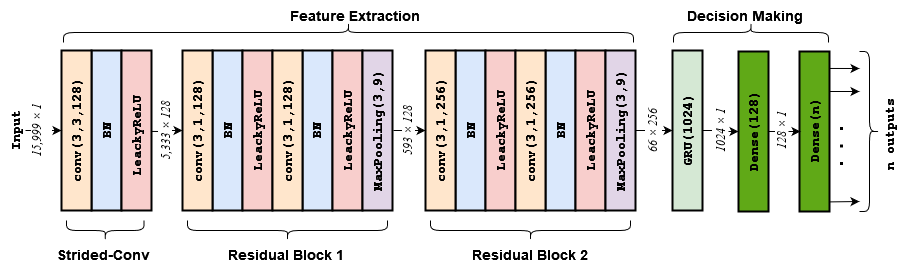} 
    \caption{Architecture of the proposed optimized shallow RawNet model. The operators utilized in the network include Conv1D(kernel size, strides, filters), MaxPool1D(pool size, strides), GRU(units), and Dense(nodes).}
    \label{fig:NNArch}
\end{figure*}

As an illustrative example, the classification problem for the data group \texttt{audio60} is denoted as $G^{\mathrm{60}}_{\mathrm{Audio}} = \{C_1, C_2, C_3\}$, where $C_1$ encompasses the audio frames from grid \textbf{A}, $C_2$ from grid \textbf{C}, and $C_3$ from grid \textbf{I}. Each sample has a label $l\in \{1, 2, \dots, n\}$.

For the last part of the flowchart in Figure~\ref{fig:flowchart}, a shallow RawNet architecture has been implemented to perform the classification. The topology of the shallow RawNet was optimized through NAS using the \texttt{Keras-Tuner} library. During this search, several parameters were fine-tuned, including the number of convolutional layers (ranging between 3 and 5), the filter sizes (128 to 256), Gated Recurrent Unit (GRU) units (512 to 1024), and dense layer units (64 to 512). After extensive experimentation, the optimal configuration for this architecture was determined to include two residual blocks. 

Specifically, as depicted in Figure~\ref{fig:NNArch}, the network begins with an input layer that processes frames of size 15,999. These frames are passed through a Strided Convolution block consisting of a one-dimensional convolution layer, batch normalization (BN), and LeakyReLU activation (with a slope of 0.01 for negative inputs). This initial block outputs a feature map of size $5333 \times 128$. The first residual block follows, comprising two convolutional layers, batch normalization, LeakyReLU activation, and a max-pooling layer, resulting in an output of $593 \times 128$. Following a similar structure, another residual block with 256 filters is applied next, reducing the output to $66 \times 256$. These residual blocks are crucial for extracting frame-level embeddings from the input data. Next, the network incorporates a GRU to aggregate these frame-level embeddings into a single ENF-level representation. The output from the GRU is then passed through a dense layer, reducing the dimensionality to a 128-dimensional vector. This layer condenses the extracted features into a more abstract, higher-level representation. Finally, the 128-dimensional vector is processed by the output layer, which uses a softmax activation function to map the vector to a probability distribution over the 9 classes, completing the classification task.

In addition to optimizing the topology of the shallow RawNet, NAS is also employed to fine-tune the hyperparameters. The optimization process explicitly targets the learning rate and parameters associated with the Adaptive Moment Estimation (Adam) optimizer \cite{kingma2014adam}. Initially, the learning rate is set within a range from $10^{-4}$ to $10^{-2}$, and the $\beta$ values for the Adam optimizer vary between 0.9 to 0.999 and 0.99 to 0.999, respectively. Following the optimization with the $\texttt{Keras-Tuner}$ library, the optimal hyperparameter settings for each data group are summarized in Table~\ref{table:hyper}. These configurations effectively balance the influence of past and current gradients, contributing to efficient optimization.

To perform grid localization using InterGridNet, we adhere to the outlined steps depicted in Figure~\ref{fig:flowchart}. After data preparation, each recording frame undergoes classification by the neural network, resulting in a probability distribution across classes as determined by the softmax activation function of the last layer. For the classification of a frame into one of the known classes, the predictions should satisfy the rule:

\begin{equation}\label{eq:Entropy}
    -\sum^n_{i=1} p_i(x) \log_2 \,p_i(x) < \alpha_1 \cdot \log_2(n),
\end{equation}
where $p_i$ is the probability for each class prediction from the softmax and $n$ is the number of classes in the frame's data group. In cases where the frame fails to meet (\ref{eq:Entropy}), it is classified as \textbf{N}. 

Subsequently, a majority voting mechanism is employed to ascertain the final estimate. The final estimate is deemed valid only if it appears in at least $\alpha_2$ of the frames' predictions; otherwise, it is designated as \textbf{N}. Through the validation process, thresholds $\alpha_1$ and $\alpha_2$ have been set to 0.8 and 0.75, respectively. This approach ensures robustness in the grid localization process by requiring a consistent majority agreement across frames for a conclusive final estimation.

\begin{figure}[t]
    \centering
    \subfloat[][Bandpass filtering is applied.]{
        \includegraphics[width=0.35\textwidth]{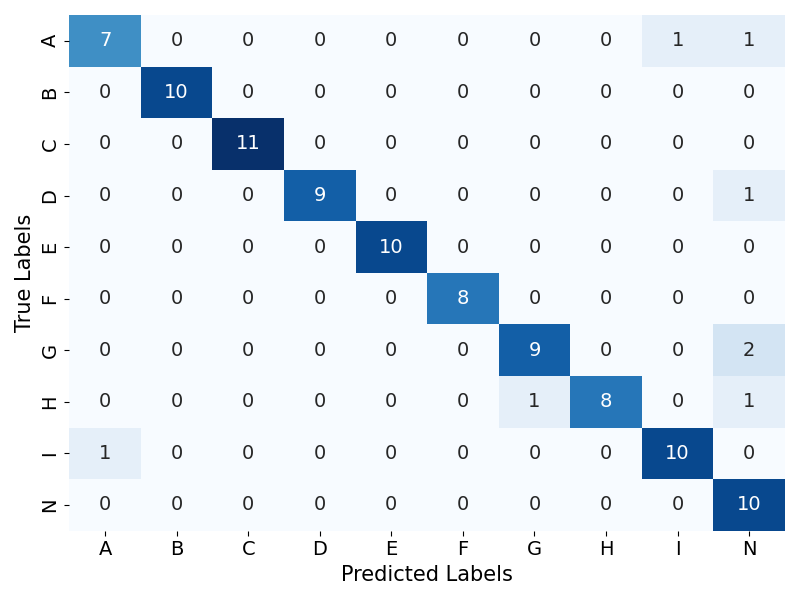}
        \label{fig:filt}
    }   
    \hfill
    \subfloat[][No filtering is applied.]{
        \includegraphics[width=0.35\textwidth]{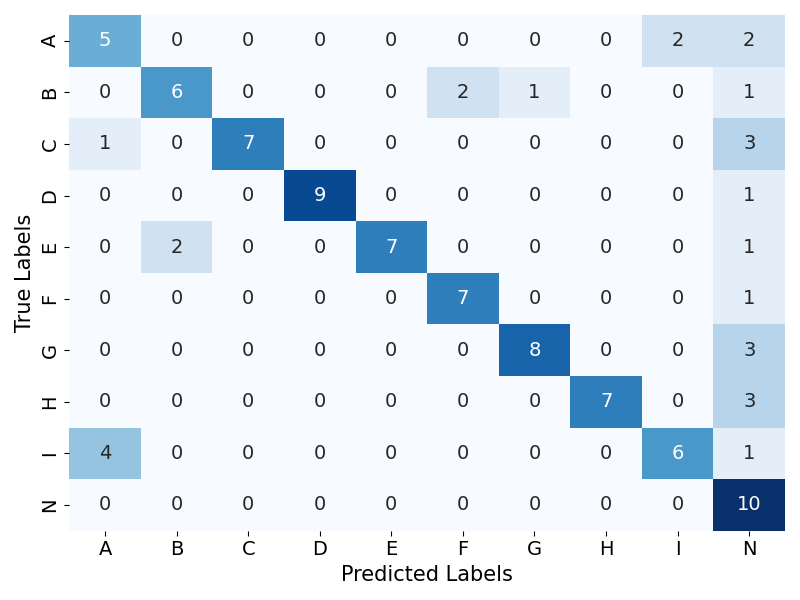}
        \label{fig:nonfilt}
    }
    \caption{Confusion matrices predictions on the testing set employing the proposed InterGridNet.}
    \label{fig:confMatrices}
\end{figure}

\section{Experimental Results}\label{sec:experiments}

In this section, the validation and testing of the InterGridNet are disclosed \cite{InterGridNet}. Additionally, limitations are discussed, providing valuable insights into the model's performance and areas for potential improvement.

\subsection{Model Validation and Testing}\label{subsec:testing}

At the training phase of each model, all available training data depicted in Figure~\ref{fig:samples}, corresponding to each data group, were utilized. For validation purposes and to experimentally determine the coefficients $\alpha_1$ and $\alpha_2$, the practice set from the SP Cup 2016 dataset was employed. This shares identical characteristics with the testing set described in Section~\ref{subsec:Dataset} and consists of 50 samples (5 samples for each class).

\begin{table}[t]
        \caption{INTERGRIDNET VALIDATION ACCURACY.}
        \resizebox{\columnwidth}{!}{
        \hfill{}
        \scalebox{0.9}{
        \begin{tabular}{*{12}{c}}
            \toprule
            \textbf{Type} & \textbf{A} & \textbf{B} & \textbf{C} & \textbf{D} & \textbf{E} & \textbf{F} & \textbf{G} & \textbf{H} & \textbf{I} & \textbf{N} & \textbf{Overall} \\
            \midrule
            Audio & 80\% & 100\% & 100\% & 100\% & 80\% & 100\% & 80\% & 80\% & 100\% & 100\% & 80\% \\
            Power & 100\% & 100\% & 100\% & 100\% & 100\% & 100\% & 80\% & 100\% & 100\% & 100\% & 96.67\% \\
            All & 80\% & 100\% & 100\% & 100\% & 80\% & 100\% & 60\% & 80\% & 100\% & 100\% & 90\% \\
            \bottomrule
        \end{tabular}}}
        \hfill{}
        \label{tb:val}
\end{table}

Table~\ref{tb:val} summarizes the accuracy achieved for each class in the practice set of applying InterGridNet after completing model training and coefficient tuning. The classifier exhibits superior performance in the Power recordings compared to the Audio recordings as the Power recordings contain stronger ENF traces, and the corresponding classifiers benefit from a larger volume of training data, contributing to enhanced performance. In addition, class ``None'' has 100\% accuracy, as shown in column \textbf{N}, underscoring the effectiveness of the ``None'' sample identification method outlined in Section \ref{Se:DescAlgo}. The aggregate accuracy of the framework culminates at 90\%.

The final assessment of InterGridNet's performance was conducted utilizing the dataset testing set. In Figure~\subref*{fig:filt}, the confusion matrix derived from the predictions of the proposed framework is illustrated, yielding an overall accuracy of 92\%. Notably, misclassifications between classes \textbf{A}-\textbf{I} are minimal, owing to the inherent constraints of the data splitting technique, which refrains from classifying a sample with ENF at 50Hz into classes \textbf{A}, \textbf{C}, or \textbf{I} with ENF at 60Hz, and vice versa. Consistent with expectations, the testing accuracy closely aligns with the validation accuracy.

\subsection{Discussion}\label{subsec:discussion}

The achieved testing accuracy of 92\% underscores the unique characteristics embedded in the ENF signal per grid. Unlike analytical feature extraction methods \cite{triantafyllopoulos2016exploring,ohib2017enf,el2005anovel,despotovic2016exploring,chow2016multi-harmonic,zhou2016geographic}, these distinctive features, crucial for solving the classification problem, are effectively extracted by the residual blocks and the GRU layer of the neural network described in Section~\ref{Se:DescAlgo}. This observation suggests that the chosen architecture demonstrates exceptional suitability for processing the ENF signal within raw audio data. 

Figures~\subref*{fig:filt} and \subref*{fig:nonfilt} present the impact of frequency filtering around the nominal ENF on the classification. When this filtering is not applied, the overall accuracy is 72\%, significantly lower compared to the scenario with bandpass filtering. This underscores the significant contribution of the ENF signal to accurately determining the grid corresponding to the recording location. In Figure~\subref*{fig:filt}, the misclassifications by InterGridNet predominantly categorize samples as ``None'' (class \textbf{N}). This exposes a vulnerability of (\ref{eq:Entropy}) in the framework but also underscores its confidence when handling samples from grids on which it has been trained. This dual observation provides insights into the framework's strengths and areas for potential improvement.

\begin{table}[t]
    \caption{TESTING ACCURACIES (\%) IN SP CUP 2016 DATASET.}
    \label{table:SpCup}
    \begin{center}
    \scalebox{0.8}{
    \begin{tabular}{p{4.5cm}cc}
        \toprule
        \textbf{Method} & \textbf{Characteristic} & \textbf{Accuracy} \\
        \midrule
        SVM \cite{triantafyllopoulos2016exploring} & One-vs-one classification & 86\% \\
        SVM \cite{ohib2017enf} & Multi-class classification & 77\% \\
        SVM \cite{zhou2016geographic} & Multi-class classification & 88\% \\
        Random Forest, SVM, AdaBoost \cite{el2005anovel} & Ensemble method & 88\% \\
        Binary SVM \cite{despotovic2016exploring} & Binary classification & 87\% \\
        Multi-Harmonic Histogram Comparison \cite{chow2016multi-harmonic} & Frequency domain analysis & 88\% \\
        \midrule
        \textbf{InterGridNet (Ours)} & Shallow RawNet & 92\% \\
        \bottomrule
    \end{tabular}}
    \end{center}
\end{table}

Table~\ref{table:SpCup} summarizes the testing accuracy of other methods using the same testing set. The data highlights the superiority of the proposed InterGridNet framework over previous works, reaffirming its effectiveness in geolocating sound recordings. Hence, InterGridNet is a powerful tool in the field, showcasing its potential for advancing state-of-the-art audio source grid location classification.

In \cite{tzolopoulos2024}, authored by our team, a fusion model comprising five machine learning classifiers was developed, trained, and tested using audio spectrograms from the nine ENF grids. This model achieved a testing accuracy of 96\%, compared to the 92\% accuracy of the proposed InterGridNet. While the higher accuracy of the fusion model can be attributed to its combination of multiple classifiers, it's important to note that it required a significantly larger parameter count, with 11 million parameters for the CNN alone, which further increased when including the parameters of the fusion framework’s classifiers. In contrast, InterGridNet, with a streamlined architecture of 7 million parameters, adopts a novel unified single-classifier approach based on raw audio input via a DNN, highlighting its innovation and efficiency in power grid classification without the need for classifier fusion.

\section{Conclusions}\label{sec:conclusions}

This paper presents InterGrid, a novel framework for geolocating audio recordings across different power grids, incorporating optimization through NAS. Inspired by RawNet's architecture, InterGridNet has employed a shallow version of RawNet, offering a dynamic framework that includes preprocessing techniques to tackle the complex challenge of inter-grid localization within audio recordings. Key techniques have been crucial, such as bandpass filtering of ENF data, integration of residual layers for extracting frame-level embeddings, and softmax activation for decision-making. This research has marked the first implementation of DNN methodology for classification with preprocessing methods, achieving a 92\% accuracy rate on the SP Cup 2016 dataset. Future research will employ a transformer architecture for grid location classification. To enhance transparency and understand the model's decision-making process, explainable AI (xAI) techniques will also be integrated to extract specific patterns associated with each grid.

\section*{Acknowledgment}
This research was supported by the Hellenic Foundation for Research and Innovation (HFRI) under the ``2nd Call for HFRI Research Projects to support Faculty Members \& Researchers" (Project Number: 3888).

\bibliographystyle{IEEEtran}
\bibliography{References}

\end{document}